\documentclass[a4paper,11pt,twoside]{article}
\usepackage{a4wide}

\newcommand{\bbint}{\int\hspace{-0.25cm}\int\hspace{-0.3cm}}

\newcommand{\eps}{{\epsilon}}
\newcommand{\om}{{\omega}}

\newcommand{\pa}{{\partial}}
\newcommand{\beq}{\begin{equation}}
\newcommand{\eeq}{\end{equation}}
\newcommand{\beqa}{\begin{eqnarray}}
\newcommand{\eeqa}{\end{eqnarray}}
\newcommand{\ben}{\begin{enumerate}}
\newcommand{\een}{\end{enumerate}}
\newcommand{\bi}{\begin{itemize}}
\newcommand{\ei}{\end{itemize}}
\newcommand{\vr}{{\vec r}}
\newcommand{\tensor}[1]{\stackrel{\leftrightarrow}{#1}}

 \title{The Aharonov-Bohm-Effect, Non-commutative Geometry, Dislocation
Theory, and Magnetism}

\author{U.\ Krey\footnote{e-mail uwe.krey@physik.uni-regensburg.de }
\\
  Inst.\ f\"ur Physik II, Universit\"at Regensburg, 93040 Regensburg,
Germany
  }
\date{Nov. 6, 2007}
\begin{document}

\maketitle
\begin{abstract}

\noindent The four items mentioned in the title are put into context in
an informal way.

\end{abstract}
{
\vglue 0.2 truecm\hrule\vglue 0.5 truecm

\section{Introduction} This is an informal paper, not intended for
publication: the items mentioned in the title

\begin{enumerate}
\item the Aharonov-Bohm effect
\item non-commutative geometry
\item dislocation theory, in particular concerning the role of the
Burgers vector \item magnetism (mainly spin-orbit interaction)
\end{enumerate} are considered and put into context. In this way we hope
to remove certain {\it high-browed} features from the issue, at the same
time clarifying the general meaning of the first and second terms, and
also putting some emphasis on the work of the community studying the
third or fourth one, often without relation to those working on one of
the two first-mentioned subjects.

\section{The Aharonov-Bohm Effect} The Aharonov-Bohm effect, see
\cite{Bohm}, is an important quantum-mechanical phenomenon showing
explicitly that quantum-mechanics is not a classical theory as usual,
for example, as Newtonian mechanics or conventional electromagnetism. A
magnetic induction $\vec B$ is considered, which gives rise to an {\it
interference effect} of electrons, which are definitely outside the
range where the Lorentz forces act and some effect could naturally be
expected. Nevertheless, a well-defined interference is observed, since
in quantum mechanics it is not $\vec B$, but the {\it magnetic vector
potential} $\vec A$ that counts; of course, the {\it closed-loop
property} of the integration path also counts, see below.

In fact, in quantum mechanics the behaviour of a charged particle  is
described by the Hamiltonian \beq {\cal H}=\frac{(\vec{\hat p} -q\vec
A)^2}{2m}+q\Phi(\vr ,t)\,. \eeq Here we work in the SI-system; $q$ is
the charge and $m$ the mass of the particle, $\vec {\hat p}$ is the
momentum operator, $\vec A(\vec r, t)$ the vector potential and
$\Phi(\vec r, t)$ the scalar potential of the electromagnetic field,
{\it i.e.}: \beq \vec B=\rm{rot\,}\vec A \, \eeq and \beq \vec
E=-\rm{grad\,}\Phi-\frac{\pa \vec A}{\pa t}\,.\eeq

In the following, for simplicity, let us consider only situations where
neither the electrical field $\vec E$ nor the scalar potential $\Phi$
plays a role.

As a consequence, classically, only the magnetic induction $\vec B$ is
important, namely through the Lorentz force, and the Newtonian equation
of motion is simply \beq m\vec a=q[\vec v\times \vec B]\,, \eeq where
$\vec v$ is the velocity and $\vec a$ the acceleration of the particle.
Usually one considers a thin magnetic wire magnetized longitudinally,
such that the magnetic induction outside the wire vanishes. But the
vector potential $\vec A$ does not vanish outside!

To keep gauge invariance, i.e. invariance of the electromagnetic fields
against changes involving an arbitrary gauge function $f$ in the form

\beq B=\rm{rot\,}\vec A+\rm{grad\,}f\,,\eeq \beq\label{eq7a}\vec
E=-\rm{grad\,}\phi-\frac{\pa f}{\pa t}\eeq it is necessary  to
concentrate on {\it closed loop} integration paths $\Gamma$, i.e. by
representing the {\it closed loop} property by the integration symbol
$\oint$ one has: \beq\label{eq7} \oint_\Gamma d\vec r\cdot\vec
A={\bbint}_{\,\,F}\,\, \vec B\cdot \vec n\,d^2A\,=:\Phi_F\,. \eeq Here
$F$ is any surface clamped into $\Gamma$ (i.e. $\Gamma$ is the boundary
to $F$, $\Gamma =\pa F$). Note that there are many different surfaces,
of which the same $\Gamma$ is the boundary; this is the genuine reason
for the gauge degree of freedom.

In (\ref{eq7}) the vector $\vec n$ denotes the {\it normal} to the
surface $F$ and $d^2A$ is the area element. The result of (\ref{eq7}) is
the magnetic induction $\Phi_F$ flux through $F$, i.e. through the wire
cross section, although $F$ can be much larger.

So, the Aharonov-Bohm effect, which was experimentally realized, e.g.,
by B\"orsch {\it et al.}, \cite{Boersch}, shows in a specific way that
quantum-mechanics is {\it more}\, than a classical theory: It is not
$\vec B$ and the corresponding Lorentz force, which counts, but the
magnetic flux $\Phi_F$ through an arbitrary closed loop $\Gamma$, or
equivalently the circumferential integral $\oint_\Gamma\,d{\vec
r}\cdot\,{\vec A}\,$, which is identical with $\Phi_F$. Thereby the
essential point is to note that the flux may correspond to a very small
part of the interior of $\Gamma$, and not to the totality of it.

\section{Non-commutative Geometry} The effect can also be interpreted as
a change of geometry induced by the magnetic flux. This is called {\it
non-commutative geometry}, see e.g. \cite{Connes}, since $[p_j,p_k]\psi
=0$, $\forall \psi$, whereas $[p_j-qA_j, p_k-qA_k]\psi=q\hbar{\rm
i}\cdot\{ [\partial_j A_k]\psi + [A_j\partial_k ]\psi \} \ne 0$ . The
term {\it non-commutative geometry}\, may look highbrowed to people from
the solid-state community, but it turns out that this is not so. In
fact, the closed line $\Gamma$ corresponds to the Burgers loop in {\it
dislocation theory}, and the magnetic induction $\vec B$ to the {\it
dislocation density} $\stackrel{\leftrightarrow}{\eta}$, a tensorial
quantity with two indexes, e.g., essentially the Burgers vector $\vec b$
of the dislocation,  times the tangent vector $\vec \tau$ of the
dislocation line. The fundamental equation between $\vec B$ and $\vec
A$, namely $B_i =\epsilon_{i,j,k}\partial_j A_k$, with the well-known
antisymmetric unit tensor $\epsilon_{i,j,k}$, corresponds to the
compatibility equation between the strain
$\stackrel{\leftrightarrow}{\varepsilon}$ and
$\stackrel{\leftrightarrow}{\eta}$, namely ${\rm Ink}\,
\stackrel{\leftrightarrow}{\varepsilon}
=\stackrel{\leftrightarrow}{\eta}$, where the symmetric incompatibility
operator, acting on a symmetric two-tensor $\xi_{j,k}$, is given by
$({\rm Ink}\, \stackrel{\leftrightarrow}{\xi})_{i,j}
:=-\epsilon_{i,k,l}\cdot \epsilon_{j,m,n}\partial_k\partial_m
\xi_{l,n}.$ This result is {\it symmetric} in the indexes i and j, and
also in l and n.

\section{Dislocation theory} The importance of  dislocation theory in
the present context has already been shown in the preceding
section. Additionally we mention the work of E. Kr\"oner, see
\cite{Kroener}, who introduced a close relation between the
source-tensor $\stackrel{\leftrightarrow}{\eta}$ of the incompatibility
and the {\it eigenstresses} of incompatible solids.

 It is well-known that in case of {\it compatibility}, the strains
$\stackrel{\leftrightarrow}{\epsilon}$ can be derived from a {\it shift
vector} $\vec u(\vec r , t)$, through the identity
$\epsilon_{i,k}=(1/2)(\partial_i u_k +\partial_k u_i)$ (for simplicity
we restrict ourselves to linear elasticity).

In contrast, in case of {\it incompatibility}, the above-mentioned
identity does not apply.  However, strains and stresses are related as
usual, and usually dislocations (Burgers vector and tangent line) are
the sources of the incompatibility, see \cite{Kroener} and
\cite{Landau}.

\section{Magnetism} Now the Maxwellian stress tensor
$\tensor{\sigma}_{Maxwell}$ comes into play (which - by the way - in a
magnetic system is not symmetric, due to the torque $dV\, \vec J \times
\vec H$, where $\vec J$ and $\vec H$ have their usual meaning, i.e.,
$dV\,\vec J$ is the magnetic moment of the volume element $dV$, see for
example \cite{Brown}).  Magnetic anisotropies, i.e., the spin-orbit
forces, are particularly important at surfaces and interfaces (i.e., the
spin-transfer across them should also be influenced)  and one should
note the effect of {\it magnetostriction}, which is often neglected,
but important for the sources of incompatibility, especially the
magnetostrictive stresses belong to the {\it non-compatible
eigenstresses} in the sense of E. Kr\"oner. If, e.g., by
magnetostriction the magnetic domains are elongated in the direction of
magnetization and compressed vertically to it, then domains with
different directions usually produce incompatible strains. Here one
should have a look at figure 14 in the above-mentioned book of Kr\"oner.

\noindent But in contrast to elastic and magnetostrictive energies,
yielding exclusively  {\it symmetric\,} stresses,
$\sigma_{i,k}=\sigma_{k,i}$, because the energy depends only on the
symmetric part of the distortions, e.g. $\eps_{i,k}=(1/2)\cdot
(u_{i,k}+u_{k,i})$, other magnetic interactions also involve the {\it
antisymmetric} part, $u_{[i,k]}=(1/2)\cdot (u_{i,k}-u_{k,i})$. This is
dual to a rotation vector $\vec\om$, e.g. $\om_3:=u_{[1,2]}$, with a
vector potential $\vec A_{\vec \om}$. This must be added to Kr\"oner's
symmetric theory, leads to {\it torsion densities} and to the appearance
of an antisymmetric part of $\sigma_{i,k}\,.$

Moreover, the Burgers loop equation $\oint_{\partial F} {\rm d}u^i\ne
0$, namely $=\int_F\,(g^{\rm dislocation})^i_{\alpha ,\beta
,\gamma}\,{\rm d}x^\beta\cdot {\rm d}x^\gamma$ corresponds exactly to
the dislocation density $(g^{\rm dislocation})^i_{\alpha ,\beta
,\gamma}=b^i\cdot t_\alpha\cdot ({\rm areal\,\,
density\,\,perpendicular}\newline{\rm to\,\,}{\rm d}x^\beta\cdot{\rm
d}x^\gamma )$, and simultanously to the equation $R^i_{\alpha ,\beta
,\gamma}v^\alpha\,{\rm d}x^\beta\cdot dx^\gamma$, with the curvature
tensor $R^i_{\alpha ,\beta ,\gamma}$ of differential-geometric spaces
Thus, dislocations, or magnetism etc., lead to curvature-like phenomena
even with trivial connection (e.g. with $g_{i,k}=\delta_{i,k}$).

\noindent All this may be well-known, but usually the phenomena are
looked upon only separately, if at all.

\section*{Acknowledgements} The author would like to thank C.H. Back, G.
Bayreuther and J. Zweck, and the members of their groups, for continuous
encouragement.


\end{document}